\documentclass[acus]{JAC2003}

\usepackage{graphicx}
\usepackage{booktabs}
\usepackage{amsmath}
\usepackage{multirow}

\begin{document}
\title{EXPERIMENT AND SIMULATIONS OF SUB-PS ELECTRON BUNCH TRAIN GENERATION AT FERMILAB PHOTOINJECTORS\thanks{The work was supported by the Fermi Research Alliance, LLC under the DOE Contract No. DE-AC02-07CH11359, and by Northern Illinois University under the DOE Contract No. DE-FG02-08ER41532.}}

\author{Y.-E Sun$^1$,  M. Church$^{1}$, P. Piot$^{1,2}$,  C. R. Prokop$^{1,2}$\\
$^1$ Fermi National Accelerator Laboratory, Batavia, IL 60510, USA \\
$^2$ Department of Physics, Northern Illinois University, DeKalb, IL 60115, USA  }

\maketitle

\begin{abstract}

Recently the generation of electron bunch trains with sub-picosecond
time structure has been experimentally demonstrated at the A0
photoinjector of Fermilab using a transverse-longitudinal phase-space
exchange beamline. The temporal profile of the bunch train can be
easily tuned to meet the requirements of the applications of modern
accelerator beams. In this paper we report the A0 bunch-train experiment and explore numerically the possible extension of this technique to shorter time scales at the Fermilab SRF Accelerator Test
Facility, a superconducting linear electron accelerator currently
under construction in the NML building.
\end{abstract}

\section{INTRODUCTION}
 Electron bunch trains with sub-picosecond structures have applications ranging from the generation of super-radiant radiation~\cite{gover,bosco, ychuang} to the resonant excitation of wakefields in beam-driven accelerating techniques~\cite{muggliprstab,jing}. Several techniques to obtain such bunch trains are developed in recent years, such as placing an interceptive mask in a dispersive section~\cite{muggli}. Our method consists introducing a horizontal beam density modulation, then exchanging the beam transverse and longitudinal phase-space, and finally recovering the density modulation in time and energy profile~\cite{yineLINAC08}.  The phase-space exchange beamline is composed of one dipole-mode RF cavity flanked by two identical doglegs, each dogleg formed by two dipole magnets (henceforth referred to as the double-dogleg beamline)~\cite{ruanPRL}. Under the thin lens approximation and phase-space exchange conditions~\cite{emma2}, the beam phase-space final longitudinal coordinates $(z,\delta)$ are related to the initial horizontal coordinates $(x_0,x'_0)$ via
\begin{equation}
 z = -\alpha x_0-\alpha S x'_0,
\delta =-\frac{1}{\alpha L}x_0-\frac{L+S}{\alpha L}x'_0,\label{eq:xzmap}
\end{equation}
where $\alpha$ is the bending angle of the dipole magnets, $L$ is the drift space between the dipoles in a dogleg, and $S$ is the drift space between the middle dipole magnets and the dipole-mode cavity. The coupling described by Eq.~(1) can be used to arbitrarily shape the temporal distribution of an electron beam~\cite{piotPRSTAB} including tunable bunch trains with sub-picosecond temporal structure,which have been experimentally demonstrated in our experiment at the A0 photoinjector~\cite{sunPRL}.
\section{EXPERIMENTAL DEMONSTRATION AT A0 PHOTOINJECTOR}
\begin{figure}[hhhhhhhhh!!!!!!!!!!!!!!]
\centering
\includegraphics[scale = 0.4]{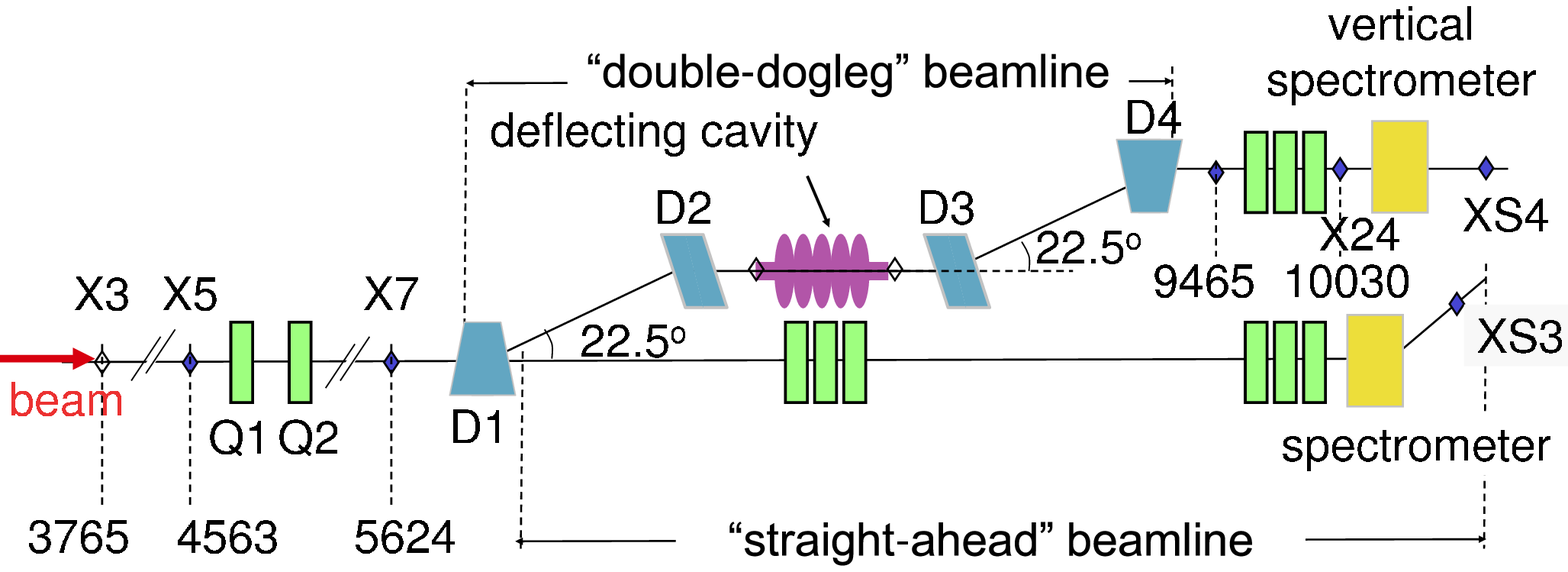}
\caption{Top view of the 14 MeV beamlines  at A0 photoinjector. The ``X" refers to diagnostic stations, ``Q" the quadrupole magnets and ``D" the dipole magnets. Distances shown are referenced to the photocathode and in the unit of mm. \label{fig:a0beamline}}
\end{figure}

The experiment of generating sub-picosecond electron bunch trains is carried out at the Fermilab A0 photoinjector. The electron beam is photo-emitted from a cesium telluride cathode, accelerated by a 1.3~GHz 1.5-cell RF gun and a 9-cell superconducting cavity, reaching an energy of 14~MeV for our experiment reported here. Downstream of the accelerating cavity, the beamline includes a set of quadrupole and steering dipole magnets, and diagnostic stations for transverse emittance measurements before splitting into two beamlines; see Fig.~\ref{fig:a0beamline}. The ``straight-ahead'' beamline is used for transverse and longitudinal beam parameter diagnostics, while the ``double-dogleg'' beamline for phase-space exchange between horizontal and longitudinal phase-spaces~\cite{ruanPRL}.

A tungsten multi-slits plate is inserted at X3 (see Fig.~\ref{fig:a0beamline}). The horizontal density of the electron beam is modulated into vertical beamlets downstream of the slits. A couple of quadrupoles (Q1 and Q2 in Fig.~\ref{fig:a0beamline}) are used to tune the transverse Courant-Synder parameters of the beam prior to the double-dogleg beamline, which provide us the controlling knobs for the time structure of the bunch train after the phase-space exchange.

Downstream of the double-dogleg beamline, an aluminium screen can be inserted at X24. Coherent transition radiation (CTR) is generated as the beam impinges on the screen. An optical path directs the CTR through the vacuum window and an Michelson autocorrelator. The intensity of the autocorrelation function is detected by a liquid helium-cooled bolometer.

\begin{figure}[hhhhhhhhh!!!!!!!!!!!!!!]
\centering
\includegraphics[scale = 0.5]{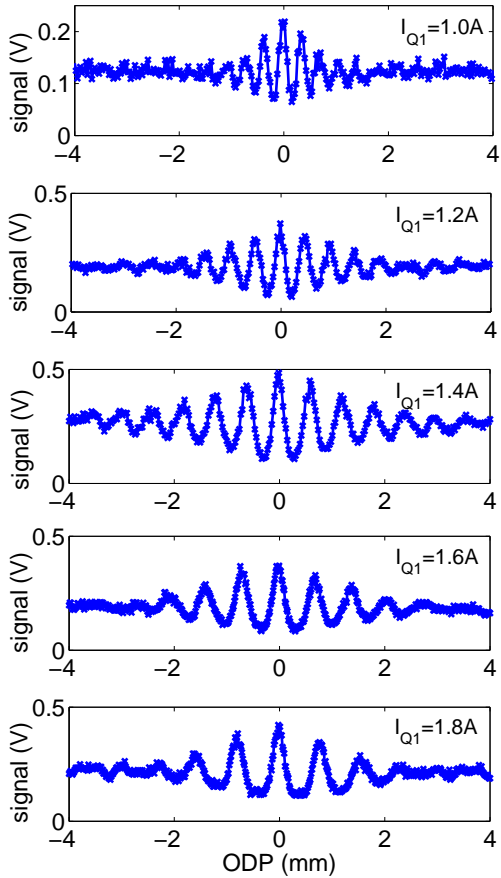}
\caption{Autocorrelation signal measured by the bolometer as the current of quadrupole Q1 is varied from 1.0A to 1.8 A. The current of quadrupole Q2 is fixed at -0.6A. \label{fig:time}}
\end{figure}

Keeping the current of quadrupole Q2 fixed at -0.6~A, we scan the current of Q1 from 1.0~A to 1.8~A. The autocorrelation functions measured by the bolometer are shown in Fig.\ref{fig:time}. The 100\% modulation of the autocorrelation function implies that the bunches within the train are completely separated. The bunch train separation is extracted from the measured autocorrelation function. The minimum measured separation is 350~$\mu$m. Assuming the bunches are Gaussian longitudinally, the estimated rms bunch length is less than 300~fs (which includes the ~200~fs resolution of our measurement system).

\begin{figure}[hhhhhhhhh!!!!!!!!!!!!!!]
\centering
\includegraphics[scale = 0.6]{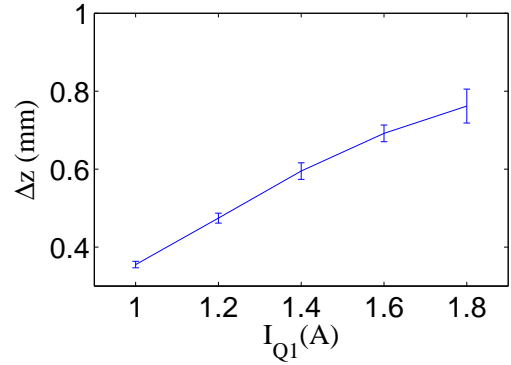}
\caption{Bunch train separation as a function of the current of quadrupole Q1. The current of quadrupole Q2 is fixed at -0.6A. \label{fig:time2}}
\end{figure}

The separation between two neighboring peaks is almost doubled as $I_{Q1}$ increased from 1.0~A to 1.8~A; see Fig.~\ref{fig:time2}. This demonstrates the tunability of the technique used here to generate sub-ps electron bunch trains.

\section{NUMERICAL SIMULATION FOR FUTURE EXPERIMENT AT THE NML PHOTOINJECTOR}
\begin{figure*}[bth]
\centering
\includegraphics[width=0.950\textwidth]{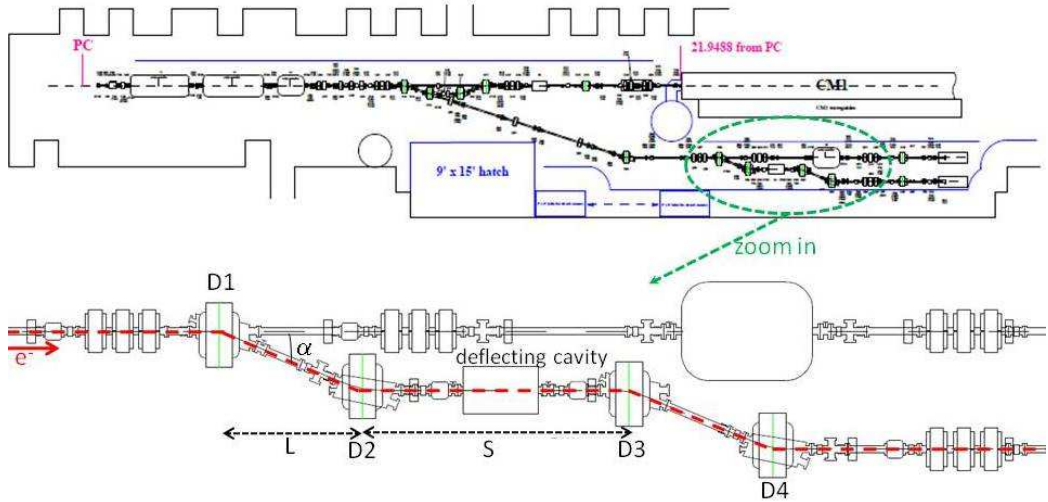}
\caption{The layout of NML photoinjector and zoom-in of the double-dogleg beam line designed for transverse-to-longitudinal phase-space exchange, ``D1"-``D4"  are the four dipole magnets. \label{fig:nmlbeamline}}
\end{figure*}

The successful demonstration of the generation of tunable sub-ps electron trains at the A0 photoinjector is very encouraging. As the A0 photoinjector is expected to cease operation at the end of the FY2011, and part of the beamline is going to be incorporated into the beamline of the Fermilab SRF Accelerator Test
Facility(ATF) , a superconducting linear electron accelerator currently under construction in the NML building~\cite{church,prokop}. The injector of the ATF (henceforth referred to as the NML photoinjector) consists of a co-axial 1.3 GHz RF photocathode gun, followed by two TESLA-type 9-cell SC RF accelerating cavities and a 3rd harmonic SC RF cavity. The beam energy of the NML injector is around 40~MeV, and the bunch charge is variable up to 3.2~nC. A transverse-to-longitudinal phase-space exchange beamline that is similar to the A0 beamline is available downstream of the injector; see Fig.~\ref{fig:nmlbeamline}. The dipole magnets in each dogleg is separated by $L$=1.07~m, and the distance $S$ between to two center dipoles is 1.99~m. Each dipole bends the beam by 22.5$^{\circ}$ while each dogleg generates a dispersion of 44~cm.

\begin{figure}[hhhhhhhhh!!!!!!!!!!!!!!]
\centering
\includegraphics[scale = 0.6]{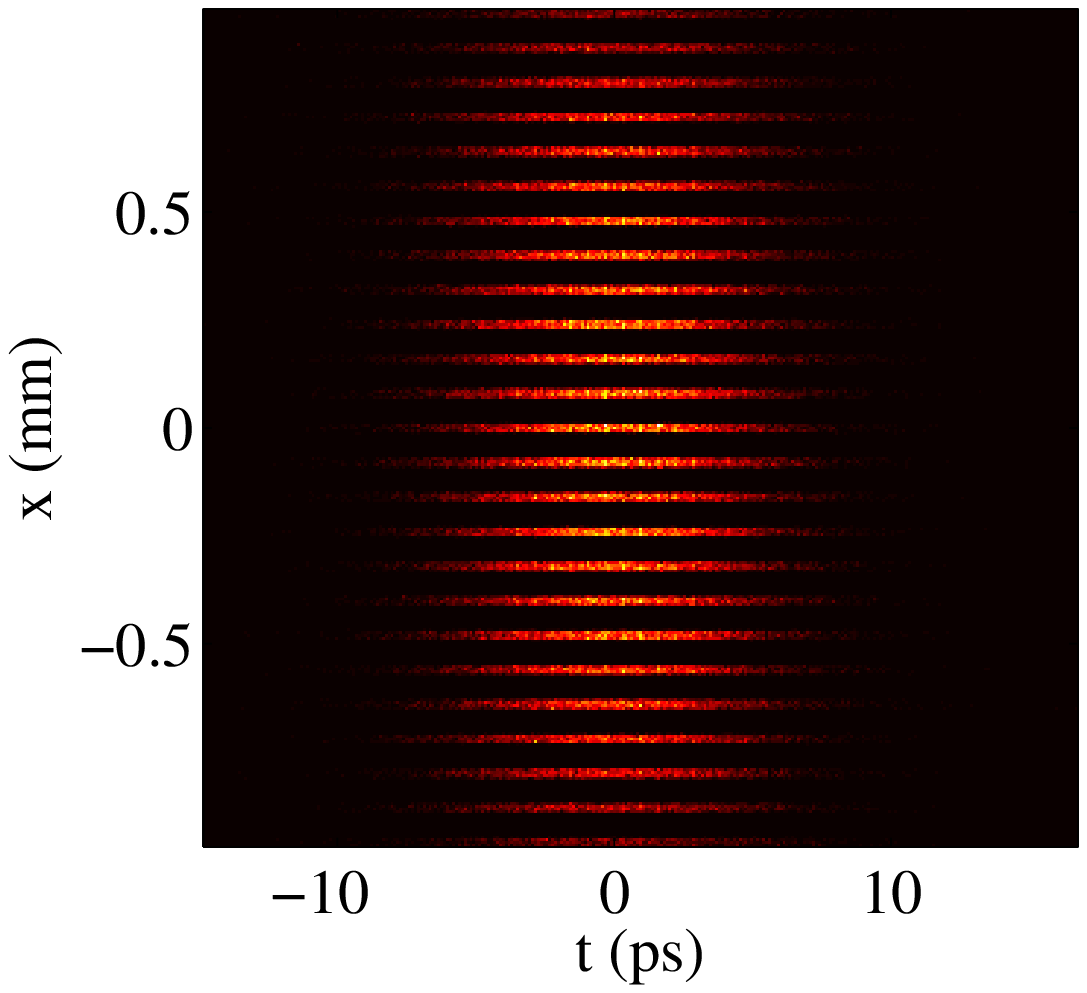}
\caption{Multibunch train generated using 20~$\mu$m-wide vertical slits separated by 80$\mu$m, 25.05\% beam passes through the slits. The bunches in the train are separated by 470~fs, each bunch rms length being 68~fs. \label{fig:xslits}}
\end{figure}

\begin{table}[h!]
\caption{\label{tab:initpara}Initial beam parameters used in GPT simulation before phase-space exchange.}
\begin{center}
\begin{tabular}{l c c c}
\hline \hline Parameter & Symbol    &   Value       & Units  \\ \hline
energy              & $E$               & 40        & MeV \\
charge              & $Q$               & 250       & pC     \\
rms duration        & $\sigma_z$        & 1.0       & mm    \\
horizontal emit.    & $\varepsilon_x^n$ & 0.6       & $\mu$m  \\
longitudinal emit.  &  $\varepsilon_z^n$& 9.7       & $\mu$m   \\
\hline \hline
\end{tabular}
\end{center}
\end{table}
The simulations are performed using GPT~\cite{gpt} with 3D space charge force on. Initial beam parameters are gathered in Table~\ref{tab:initpara}. A mask with vertical slits of width $w$ and separation $d$ intercepts the beam upstream of the quadrupole triplet in front of the double-dogleg beamline (see Fig.~\ref{fig:nmlbeamline}).  The beam horizontal density is modulated by the slits, as seen in the snapshot of $x$-$t$ configuration-space shown in Fig.~\ref{fig:xslits}, where $x$ is the  horizontal position and $t$ the time within the beam. While the bunch charge is 250~pC before the slits, only 25\% of the electrons are transmitted through the slits for the two slit-configurations reported in this paper (30~$~\mu$m-wide slits separated by 120~$\mu$m, and 20~$\mu$m-wide slits separated by 80$~\mu$m).
\begin{figure}[hhhhhhhhh!!!!!!!!!!!!!!]
\centering
\includegraphics[scale = 0.6]{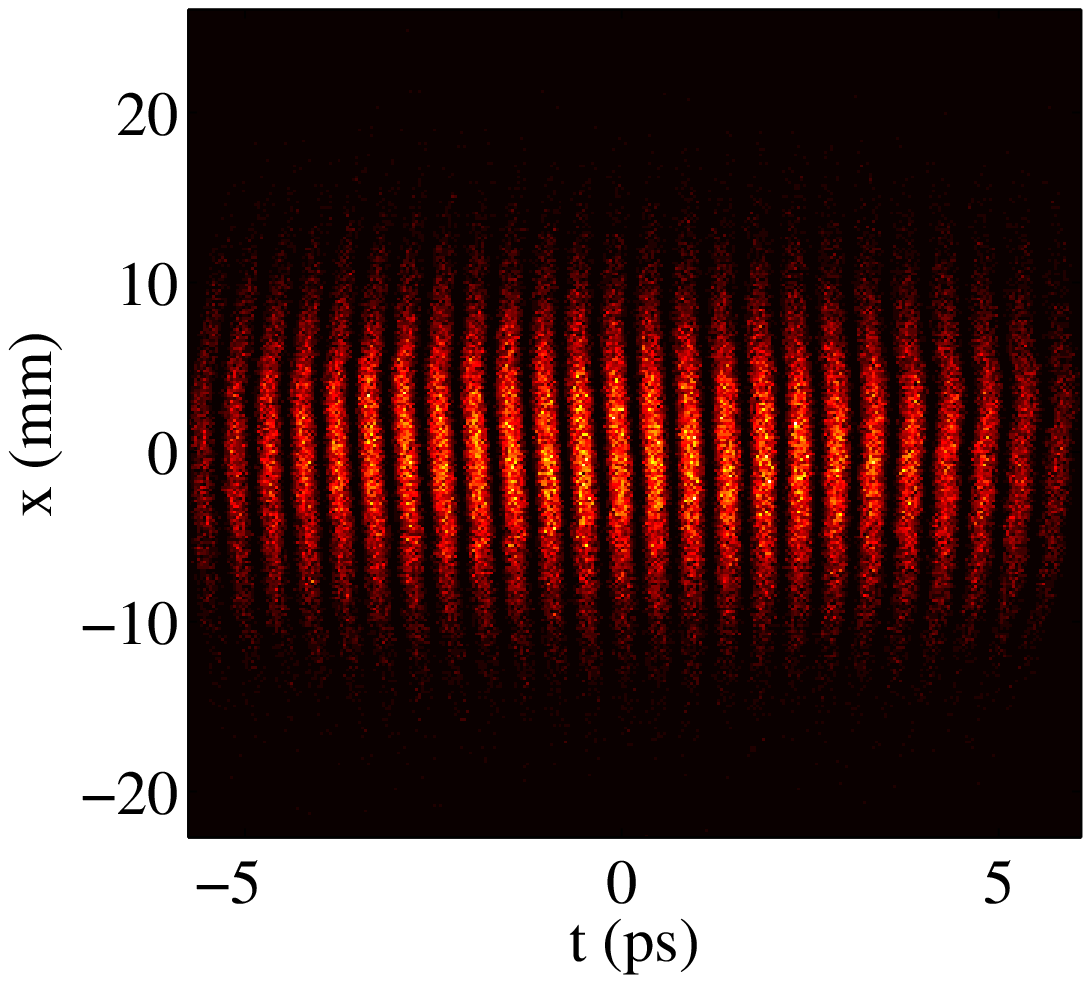}
\caption{Multibunch train generated using 20~$\mu$m-wide vertical slits separated by 80$\mu$m, 25.05\% beam passes through the slits. The bunches in the train are separated by 470~fs, each bunch rms length being 68~fs. \label{fig:zslits}}
\end{figure}

The optimization parameters for phase-space exchange and bunch train generation include the gradient and phase of the deflecting cavity, initial beam energy chirp and transverse Courant-Synder parameters upstream of the double-dogleg beamline. The bunch train temporal structure can be tuned by using different slit configurations and/or varing the quadrupole strengths prior to phase-space exchange. The initial horizontal modulation is mapped into longitudinal coordinates (time and energy) via Eq.~(\ref{eq:xzmap}). In Fig.~\ref{fig:zslits},  we see that the initial modulations existed in  $x$-axis is mapped into $t$-axis, thus a bunch train with sub-picosecond temporal structure is generated.

\begin{figure}[hhhhhhhhh!!!!!!!!!!!!!!]
\centering
\includegraphics[scale = 0.6]{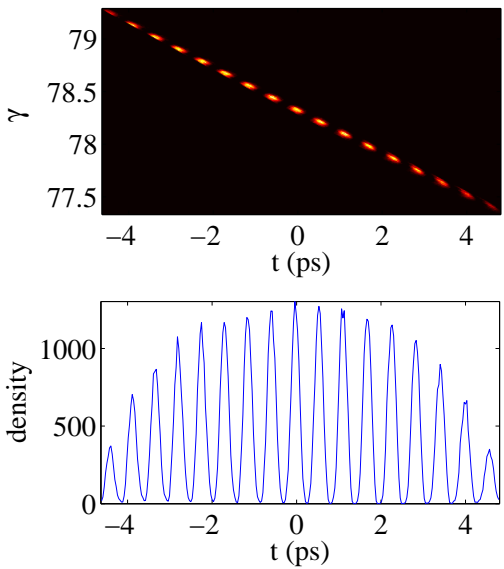}
\caption{Multibunch train generated using 30~$\mu$m-wide vertical slits separated by 120$\mu$m. The bunches in the train are separated by 560~fs, each bunch rms length being 85~fs. \label{fig:simu1}}
\end{figure}

\begin{figure}[hhhhhhhhh!!!!!!!!!!!!!!]
\centering
\includegraphics[scale = 0.6]{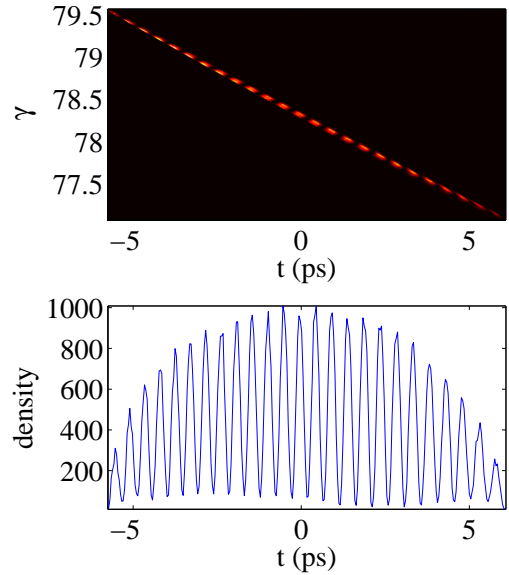}
\caption{Multibunch train generated using 20~$\mu$m-wide vertical slits separated by 80$\mu$m. The bunches in the train are separated by 470~fs, each bunch rms length being 68~fs. \label{fig:simu2}}
\end{figure}

Given the double-dog beamline as designed in the NML photoinjector, bunch trains separated by 560~fs and 470~fs are achieved using slits separated by 120$\mu$m and 80$\mu$m, respectively; see Fig.~\ref{fig:simu1} and Fig.~\ref{fig:simu2}. In each case, the quadrupoles upstream are tuned to obtain completely separated bunches in the train with minimal separations. Compared with the experimental demonstration at A0 photoinjector, the bunch train separations obtained in these simulations are reduced by a factor of 2. The advantage of the NML photoinjector beam is its smaller initial horizontal emittance, but a disadvantage is that distance $S$ between the two center dipoles is 1/3 longer than the case of A0. As the initial beam is only modulated in $x$-position and not in the $x'$ divergence, and contribution of the divergence of $x'$ to the final $t$ is linearly proportional to $S$~\cite{yineLINAC08}, a larger $S$ means more dilution of the bunch train structure at the end of the phase-space exchange.
\section{SUMMARY AND OUTLOOK}
We have experimentally demonstrated electron bunch trains separated by as low as 350~$\mu$m at the A0 photoinjector. Using a vertical multi-slit plate with smaller slit width and separation, more closely spaced bunch trains can be achieved in numerical simulations, as better beam emittance at 250~pC is expected at the NML injector. The fabrication of such multi-slit plates might be a challenge, but reducing the separation between the two center dipole magnets in the NML beamline can relieve the pressure on the requirement of slit separations. Future research on slits fabrication and double-dogleg beamline arrangement will be pursued.
\section{ACKNOWLEDGMENTS}
We thank H. Edwards, E. Harms, A. Johnson,  E. Lopez, R. Montiel, W. Muranyi, A. Lumpkin, T. Maxwell, J. Ruan, J. Santucci, C. Tan, B. Tennis, J. Thangaraj and R. Thurman-Keup for their support on the experiment.

\end{document}